# Internet of Things: System Reference Architecture

Milan Milenkovic

IoTsense LLC, Dublin, CA, USA

milan@iotsense.com

**Abstract**

This document describes an IoT system reference architecture. It is a general system architecture in the sense that it embodies key system components, functions, and flows that are commonly encountered in IoT systems.

Much of the existing literature on IoT system architecture covers either specific commercial solutions or abstract architecture descriptions that are conceptual but not readily reducible to practice. This paper describes a general, vendor-neutral architecture of the IoT system derived by abstracting the common requirements and features of a variety of system specifications and implementations. It covers architectural and design principles for constructing the core IoT system overlay, specifically functions and components involved in acquiring data, real-time processing, storage and delivery to applications and services, such as visualization and AI. The appendix outlines the principles of IoT information modeling and data and metadata handling. As a reference architecture, this document is meant to serve as a template and a starting blueprint for constructing specific IoT system solutions.

The intent is to maintain this material as a living document and evolve it over time into a broader technical community consensus by incorporating and acknowledging the accumulated feedback and contributions. Direct comments and suggestions are invited now. More structured mechanisms of managing crowdsourcing and community collaboration are planned for implementation if warranted.

## 1 INTRODUCTION AND PURPOSE

This document describes a reference IoT system architecture. It is a general system architecture in the sense that it embodies key system components, functions, and flows that are commonly encountered in IoT systems. As a reference architecture, it is meant to serve as a template and a starting blueprint for constructing specific IoT system solutions.

Much of the existing literature on IoT system architecture covers either specific commercial solutions [1, 2, 4] or abstract, often layered architecture descriptions [3, 8, 10, 17] that are conceptual but not readily reducible to practice. This paper describes a general, vendor-neutral architecture of the IoT systems functional overlay derived by abstracting the common requirements and features of a variety of system specifications and implementations.

The primary audience for this document are IoT system architects, designers, developers and technical decision makers who are building or evaluating IoT solutions.

This document covers IoT system functions and components involved in acquiring data, real-time processing, storage and delivery to applications and services, such as visualization and AI. These are core IoT-specific components that can be viewed as a functional system overlay that spans the range from sensor data acquisition on one end to applications and services that operate on aggregated data on the other end. The document describes the architectural principles and practices for its construction. Details of sensor operation and interfacing, as well as design of insight-producing applications and services, such as IoT analytics and ML/AI systems, are outside of the present scope and they are described elsewhere [12].

Principles of the physical thing information and data modeling necessary for achieving M2M semantic interoperability are described in the appendix. M2M semantic interoperability s a requirement in IoT systems that designers need to address since no suitable Internet technology to solve it exists.

The document is organized as follows. Section 2 covers IoT system purpose and a functional view of major components that carry it out – data collection, forming of insights, and acting upon them. Section 3 provides on overview of IoT system infrastructure and identifies key components that implement data and control planes. Section 4 provides a more detailed description of edge functionality. Section 5 describes the cloud portion of IoT core system functions and components, including data ingestion, digital twins, real-time stream processing, and sensor database storage. Section 6 describes components of the control plane, security and management. Section 7 is a condensed summary of IoT system implementation considerations when designing specific solutions. The appendix outlines the principles of IoT information modeling and data and metadata handling.

## 2  IOT SYSTEM FUNCTIONAL VIEW

The primary purpose of an IoT system is to collect real-world status data and make them available to services and applications that create insights and act upon them by affecting the physical system under observation in some way. Implementation of those functions requires an infrastructure to run them and control functions to keep the IoT system secure and operational.

Figure 1 depicts a highly abstracted functional view of an IoT system with focus on data flows and types of processing from capture to output actions. It highlights three key steps in IoT data and control flows (1) data collection, (2) processing, and (3) acting upon the world based on the outcomes.

Collected data quantify the state of the physical world and various types of processing provide insights and determine the nature of actions to be undertaken when appropriate. This general description is applicable to almost any automated control system. Some of the differences made possible by IoT systems include the broad scope and variety of potential data sources, global reach by virtue of Internet connectivity, and the ability to combine and aggregate data across application domains and geographical locations. This in turn facilitates creation of services that can generate insights, optimizations and predictions based on an unprecedented diversity and scale of data.

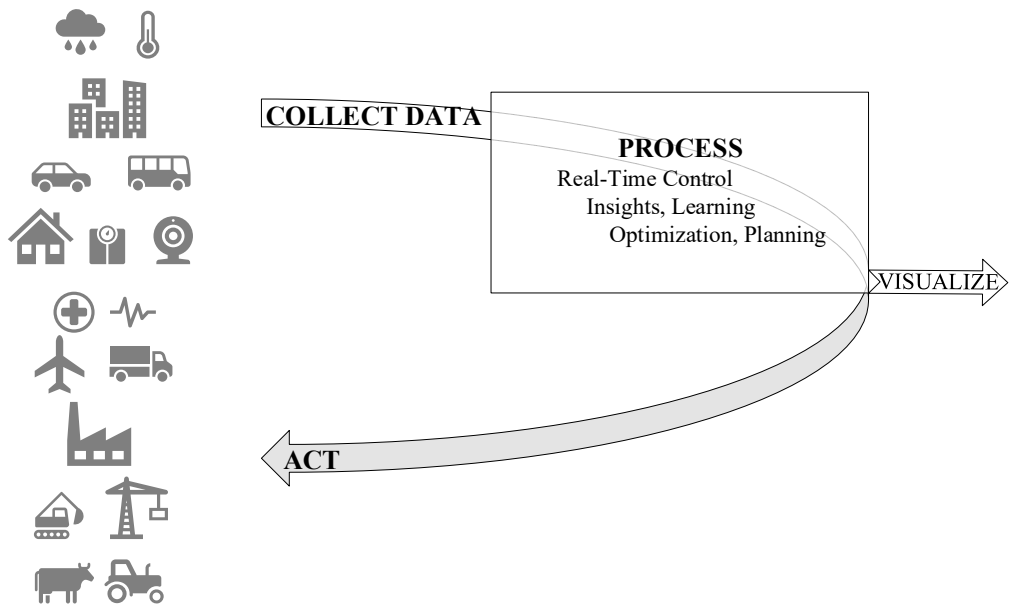

Figure 1 IoT System Functional View

## 2.1 Data Collection

Data collection starts at the edge, with a sensor acting as a physical-cyber interface that monitors and reports states of some physical entity or device. This implies the existence of real-world instrumentation via sensors and an implementation of appropriate control points via actuators. The intent is to produce a digital representation suitable for use in the cyber space. This process may involve many implementation details, such as signal conditioning, analog-to-digital conversion, scaling and conversion to engineering units [12]. From the IoT system functional point of view, data collection produces digitized state samples of the physical world suitable for processing by the applications and services in the cyber domain.

## 2.2 Data Processing

Collected data may be processed as they arrive, stored for subsequent reference and analysis, or both. Depending on the age (time stamp) of data retrieved by an application, they are often referred to as data in motion and data at rest. Data-acquisition modules usually perform the preliminary analysis and filtering to determine what to do with each particular data item or stream. Based on data values, source and system processing rules in effect, data may be routed directly to applications, stored, or discarded.

Common data pre-processing steps may include filtering such as comparison against configured thresholds to detect if the sampled data is in a special condition that warrants additional action, such as creation of an event or notification. In addition, notifications and alarms may be configured to be sent to responsible personnel as emails or texts.

Various forms of data processing may be implemented in increments or in entirety in different components of IoT systems. In general, their scope and complexity tend to increase in higher levels of system hierarchy, where more processing, power, storage and larger aggregations of data are available. Types of processing range from simple control-loop algorithms performed on incoming streaming data as they arrive, to sophisticated forms of analytics and machine-learning algorithms that operate on combinations of streaming and archived data, events, and records of past behaviors and observations of the system.

More elaborate and arguably useful levels of data processing in IoT systems involve machine-assisted analyses and optimizations. They include formulation of insights such as detecting or predicting imminent component failures by continuously comparing the observed states to a known normal operational behavior. That behavior can be established from manufacturer's specifications and design documents, system models, and past history of recorded comparable system states and outcomes of control actions.

The next level of sophistication in data processing is to provide optimizations and predictions of system behavior based on its current state, past behaviors, and guidance form algorithms, such as analytics and machine-learning models. In a more complex hierarchical system, analytics may be performed in distributed fashion at different levels. In such systems, algorithms and procedures at the lower layers may be informed and enhanced by algorithms and settings from the larger-scope cloud analytics that works on large data sets to extrapolate more general trends. This allows improvements resulting from global insights to be applied to similar or related components anywhere in the system.

## 2.3 Acting

Acting upon insights and predictions is the output and the ultimate purpose of deploying IoT systems. Actions can take different forms, from the simple remote actuation operator commands in response to status indications on the system dashboard, to automated guidance of control points that proactively manages conditions in a smart building in a manner that maximizes user comfort and optimizes energy efficiency. Actions can be implemented as direct actuation or indirectly, in the form of advice to system operators or optimizations resulting in adjustments to operational settings of the system, such as a building or a manufacturing process. They can also include identification of causes of failures and anomalous conditions followed by direct or indirect execution of the appropriate remediation actions.

Insights and resulting actions can operate in rather complex systems and domains such as smart buildings that can have tens of thousands of sensor and actuation points. With the addition of interoperable data formats and Internet connectivity, the scope of data aggregation in IoT systems can grow to potentially any level - from a single-building domain to multi-domain systems such as smart cities and regions. In theory, one could aggregate behavioral and energy efficiency data from all IoT managed buildings in a large region and use it to improve AI and train ML algorithms for their optimization. In turn, those results could be applied to a vast range of buildings or even globally, and continuously improved with influx of new data and analysis of the results of previous actions.

## 3 IOT SYSTEM: INFRASTRUCTURE OVERVIEW

This section provides an overview of IoT system infrastructure - data processing, storage, and communication components that host and execute its functions outlined in the previous section – data collection, processing, and action. This document focuses on the core IoT system components involved in gathering, pre-processing and storing of data for use by applications and services that generate insights. Their simplified view is presented in Figure 2. Core IoT functions at the edge and in the cloud are described in detail in later sections. They interface to and feed the (usually cloud based) IoT applications and services that generate insights, e.g. visualization for human insights and ML, AI for machine-generated insights. They can be combined with other business systems to provide holistic insights and generate actions that optimize business objectives or whatever the intended purpose of the IoT system may be.

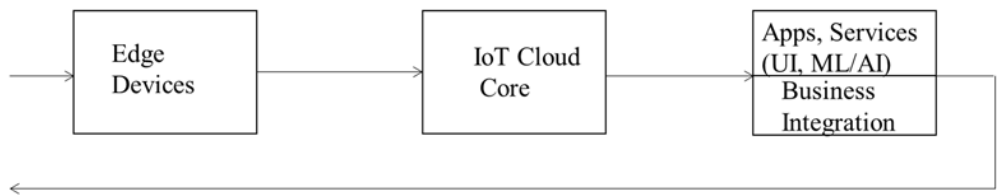

Figure 2 Major IoT System Components

Large IoT installations can be complex distributed systems with a variety of components and multiple levels of hierarchy.

A more detailed structural and hierarchical view of an IoT system and some of its key components is depicted in Figure 3. Edge components are depicted towards the bottom, communications layer mostly in the middle, and upper levels of system hierarchy ending with the cloud are shown on top.

Three major activities that need to be mapped to the infrastructure of an IoT system may be summarized as:

- Data acquisition: conversion to digital and pre-processing
- Communication: data and metadata serialization and transport
- Data processing, storage, and aggregation: repeated at multiple stages in the system hierarchy, progressively increasing scale

Edge components include sensors, smart things, gateways and fog nodes. The communication layer provides connectivity among system components that they can use for horizontal peer-to-peer interactions within a level of system hierarchy or for cross-level communication towards the Internet and the cloud. The cloud portion contains the back-end part of the IoT infrastructure where large-scale data aggregation and processing take place.

### 3.1 Data Acquisition and Edge Components

Depiction of the edge layer in Figure 3 includes several types of nodes, such as smart things and gateways, in some typical configurations. It shows a smart thing with a sensor and actuator that connects directly to the cloud via Internet, and several other nodes attached to the communications layer.

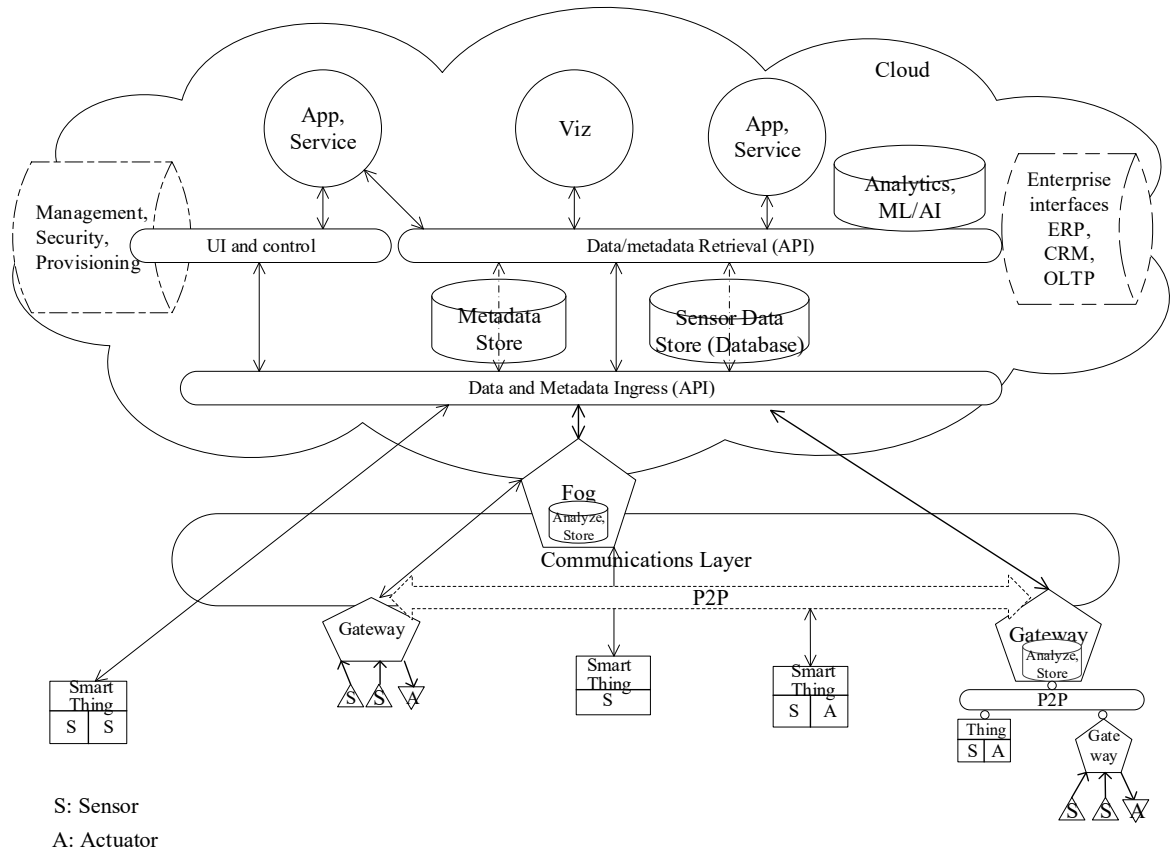

Figure 3 IoT System Infrastructure View

Gateways are edge devices in IoT systems to which one or more basic sensors are connected and dependent upon for wide-area connectivity and optional additional services. Their basic role, as the name implies, is to provide connectivity between locally connected sensors and the Internet.

Gateways can reside in physical structures, such as a building, where an Internet access point may be available and the gateway can be connected simply via a LAN connection. If a gateway is not near a usable existing Internet access point, depending on the distance to such a point, it may provide a medium-range link, such as LoRa, or a wide-area link to the Internet, via a private wide-area network or a telephone company link such as LTE or NB-IoT.

## 3.2 Communications Layer

In IoT systems, the communications layer may include a variety of wireless and wired links, spanning local areas and including long-haul connections, such as wide-area networks and telco IoT variants. It may represent a complex infrastructure of links, bridges and routers that can transport payloads from local point-to-point segments all the way to any endpoint and application on the Internet. In addition to hardware, this requires implementation of a number of network layers and protocols which in IoT systems are commonly based on the Internet layered network design blueprint [7].

The communications layer enables a vast array of edge devices and things to exchange messages with each other, the rest of the IoT system, and ultimately the Internet. At the edge and in specific domains, such as the industrial control, some proprietary and non-Internet compatible protocols may be used to connect nodes. However, to qualify as IoT, at some point in the system a transition to Internet compatible connectivity and protocols needs to be made in both directions, so that functional exchange of messages can take place with the authorized endpoints anywhere on the Internet.

In practice, communication is carried at the behest of software agents that implement the relevant functions and services at the sending and the receiving party. If both parties reside on the same network or are at the comparable levels in system hierarchy, this communication is referred to as peer-to-peer (P2P). P2P communications are sometimes also referred to as machine-to-machine (M2M) communications. In general, all communications between IoT devices themselves as well as with the applications and services are of the M2M type, so we use the term P2P to depict peer level communications, which are shown as horizontal paths in Figure 3, as opposed to the edge to cloud which would follow the more vertical paths.

P2P communications can be somewhat distinct in the sense that they can take place between nodes using simplified or application-specific protocols that are not Internet compatible. This may be done to support legacy devices or to reduce the load and simplify the design of constrained nodes. For example, devices in a home may use Zigbee or Bluetooth to communicate within a local enclave and to coordinate their group behaviors and actions. Similarly, edge sensors and simple logic controllers in a manufacturing setting may use one of the protocols for industrial automation, such as MODBUS. However, at some level, full-function IoT systems are expected to support Internet-style connectivity and communications protocols. This needs to be taken into account in the design of the communications part of IoT systems.

For example, Figure 3 illustrates some representative types of IoT connectivity configurations, such as smart things connecting directly to the cloud, or to the local network for P2P communications or as an intermediary to cloud connections. Depending on the underlying physical network, a single node may be able to engage in both types of connections.

## 3.3 Cloud

IoT data from multiple end points, gateways, and even domains are aggregated in the cloud and funneled to the system-wide applications and services. Streaming and archived data are made available to authorized applications and services via APIs and queries. Cloud implementations generally provide the top-end data aggregations, and the server, storage and connectivity infrastructure to execute services and applications that use them.

IoT data coming to the cloud may be processed in-flight as they arrive, stored for subsequent processing, or both. System-level rules and policies in effect determine how individual data should be processed and how to route them to the appropriate destination.

The cloud portion of an IoT system also provides integration and interfacing opportunity with other enterprise systems, such as Online Transaction Processing (OLTP) systems for commerce and billing, Enterprise Resource Planning (ERP) systems for holistic business insights and management, and Customer Relationship Management (CRM) systems for customer support.

Figure 3 also highlights two important interface layers in an IoT system, notably the data and metadata APIs between the edge components and the cloud, and the data-retrieval APIs for live streams and stored data used by the applications and services executing in the cloud. In general, APIs should enable applications to query, search, and access data and metadata of interest, as well as issue actuation commands. Formalizing the types of interactions and data formats that they support is not only a good design practice, it also provides the foundation for modularity and interoperability in the implementation of IoT systems.

## 3.4 Control Plane

Components and functions described so far implement the primary objective of an IoT system to collect, process, and act on data. Parts of the system that carry out these production activities and implement related system flows are commonly referred to as the data plane or user plane. The task of keeping the IoT infrastructure itself running and secure is usually delegated to a sperate system overlay that is referred to as the control plane. It is partially depicted in Figure 3 as the service and database labeled Security, Management and Provisioning. Although seemingly playing a supporting role to the primary mission, security and management are essential to keeping an IoT system up and running securely and with integrity so that it can fulfill its intended purpose and not be a threat to safety of the people and the environment.

During the normal operation, control-plane systems constantly monitor activities that may impact security and availability. This is accomplished through the network of management agents that are installed on nodes and system components to observe and report status and changes. Their reports are customarily aggregated and visualized to operators at a central control point. The agents are also used to distribute and manage security policies and credentials, change configuration, and update firmware and software as necessary. Communications required to complete these actions usually need to be protected and sequestered either by being implemented as a separate overlay, such as a Virtual Private Network (VPN), or as a separate control network.

An important function of central monitoring is the detection and analysis of suspicious behaviors that may indicate probes from the adversaries or breaches of security. When incidents are detected, the handling mechanisms and policies are activated to mitigate the situation by identifying and isolating the compromised parts of the system. Details of those operations are presented in section 6.

While the system is in operation, new nodes may need to be added, existing ones patched, and old ones decommissioned without bringing down other parts of the system or relaxing its security posture. Security and management systems are involved in preparing nodes for joining the system in the early stages of their lifecycle that precede activation. During the process of node commissioning and provisioning that follows installation, they are issued system identities and security credentials necessary for authentication and secure operation upon their activation in the system. During that time, nodes are also entered into device registries and other backend systems that may need to be involved in their operation, such as billing, asset management, and support.

## 4 EDGE NODE FUNCTIONALITY

IoT edge node and thing implementations provide a wide spectrum of functionality, from the minimalistic data acquisition and transfer in integrated sensors, to sophisticated data processing, storage and analytics in high-end gateways and fog nodes. Most of them provide a similar set of functions such as acquisition, processing and transmission of sensor data. To simplify the exposition, we describe edge functions using IoT gateway as the representative node and description vehicle given that it can embody all of those functions and perform them on behalf of multiple endpoints in its care.

An IoT gateway links sensors and things at the edge with higher levels of system processing hierarchy and the cloud. It is commonly a point where Internet connectivity is achieved. The gateway is also a security boundary between things with varying levels of security and the secured IoT processing IT infrastructure. By virtue of its placement and function, an IoT gateway can also be an interface and a boundary between production- and process-level Operational Technology (OT), such as PLCs, and the IoT Information Technology (IT) part where most of the advanced data processing and storage takes place.

As its name implies, the basic and traditional function of an IoT gateway is to act as a communications bridge between sensors and actuators at one end and the Internet and cloud on the other. This requires incorporation of appropriate hardware interfaces and processing logic for sensor data acquisition and actuation as well as protocol-converter engines for "south" side wired or wireless communication links that it may support. On the "north" side, it means hardware interfaces for the types of uplinks that are supported, ranging from Wi-Fi and Ethernet to long-haul communications such as telco lines with the appropriate modem circuitry. Uplink communication is usually Internet compatible at some point, so gateways and smart things need to include the appropriate protocol stacks for TCP/IP and HTTP often with support for transport-layer security, TLS.

The primary function of an IoT edge node is to perform or assist data collection and to provide the connectivity and security for transfer of sensor data to other system components. As indicated earlier, data may be acquired directly from sensors and transducers or from things with digitized data sources via wired or wireless links that may or may not be IP-compatible.

Common data-plane functions of a complete functional edge node, such as a gateway or a smart thing, may be categorized as:

- Core functions: data acquisition, transmission and actuation
- Optional functions: data storage, event & alert processing, control (automation)
- Advanced functions: analytics

In addition, the gateways often host local agents to support a system wide implementation of control-plane functions, including security and management.

In the sections that follow we describe major data-plane functions of edge IoT components.

## 4.1 Data Plane Functions

Sensor data are acquired at the edge by smart sensors and simpler sensors connected to gateways. Processing rules and policies in effect determine the relative significance and ultimate destination of the data. Consequently, some data samples may never leave the edge and instead be processed, stored locally, or discarded. Data and events that are destined for additional processing elsewhere in the IoT system may be communicated to peer nodes and to applications and services anywhere else in the system.

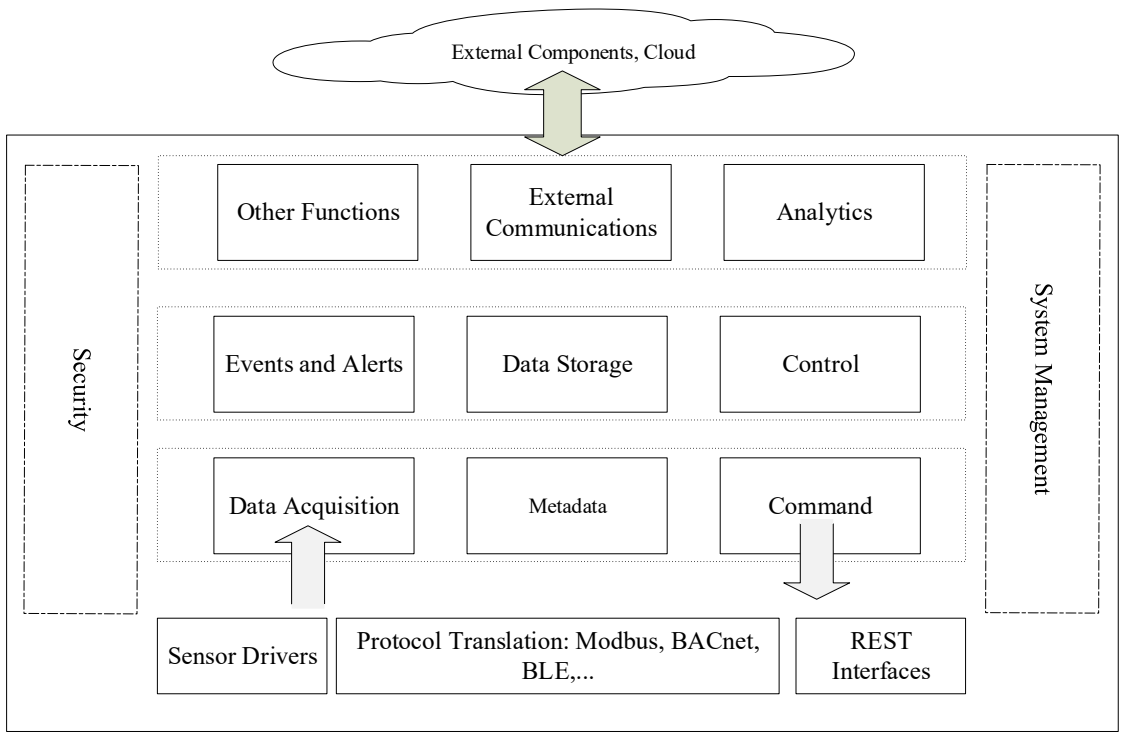

Figure 4 Edge Node Functional Components

*4.1.1 Data Acquisition*

Data acquisition refers to sampling of sensor inputs at rates that may be fixed or selectable via configuration settings and changes. Time stamp of data capture may be recorded and associated with data items at this point. As discussed earlier, sensor data need to be linearized and converted to values useful for processing and exchanges with the rest of the system, such as common engineering units of measurement. Gateway data-acquisition modules need to be able to complete whatever sensor-specific processing steps may be required. In Figure 4 sensor driver function is in charge of processing data from raw and non-standard sensors and it may include sensor-specific driver modules for that purpose.

Some of them are depicted in Figure 4, labeled the protocol translation box. An IoT gateway needs to support protocols used by device types that are connected to it. Protocol translation means that the gateway needs to be able to parse and interpret received data packets and convert them to the appropriate common internal format used by the data acquisition module. On their output from the gateway, protocol translators need to convert commands from the common gateway data information model to the format used by the target recipient, such as a PLC or an actuator. In addition to actuation, output commands may include controls and configuration, such as instructing a PLC to report its data or setting the sample rate for its sensors.

IoT data from individual sources are successive samples of sensor values and are reported as time-series data, also referred to as data channels. Their capture time is marked by a time stamp either by the originating entity or by the gateway upon entering the data acquisition module. Data and their associated time stamps are then funneled to other functional modules such as external communications, data storage, or events and alerts. Implementing these functions as microservices with well-defined APIs allow flexible configuring of data pipelines to meet the requirements of the target system application.

The functional box labeled REST is a generic representation of interface to things and gateways in the lower layers of system hierarchy. They may use their own data formats and functions and expose them via REST APIs over the compatible links and protocols.

*4.1.2 External Data Communication*

The external communications module in Figure 4 is in charge of input and output communications with the external parties, such as other peer nodes at the edge, fog and cloud levels. Unless the same information model or framework is used at both ends of the transfer, it needs to translate the outgoing data into the format agreed upon with the receiving end. Message serialization (for outgoing) and deserialization (for incoming) is also commonly performed at this point. Serialization defines how individual bits and bytes appear in actual transfers over communication links, colloquially referred to as "on the wire".

Inputs received by the gateway can include a variety of items, including data and state from peer nodes, commands for actuation, and policies and settings for various functions. These may include event-policy updates, data sampling and retention settings, control script changes, model or algorithm updates for analytics. External requests and APIs to gateway functionality may be parsed at this point or routed to the appropriate functional module unless they expose their own access points directly. Loading and updating of software modules and security updates are generally handled as a part of the control-plane functionality described later and they may use separate communication paths and credentials for that purpose.

In addition to the core functions related to data acquisition and transfers, edge nodes may perform other optional functions, such as data filtering and event or alert generation, local storage, local control and front-end analytics. They are described in more detail in sections that follow.

*4.1.3 Events and Notifications*

Events and notifications functions process incoming sensor data to detect conditions and state changes that require special handling. While there is no generally agreed upon definition, events may be regarded as any observable occurrences in the system while alerts are the more significant or critical events that may need immediate attention. State changes of binary nature, such as a valve being opened or closed, may be treated as events or simply preconfigured to always be reported to the cloud when detected. Alert notifications, when required, may be sent as notification messages from the gateway or forwarded for processing to a higher tier that aggregates all such messages and provides a notification service to specified recipients.

*4.1.4 Local Control*

Local control capability is commonly provided in systems where latency is critical and/or edge autonomy is required. Low latency is ensured by the local execution of control due to its network proximity to data sources. Autonomy refers to the requirement for some control functionality to remain operational even during the intervals when the node is disconnected from the rest of the system.

Control, when implemented locally, enables the edge node to execute pre-defined control sequences that generally cause local action when a specified condition occurs on some combination of local data.

*4.1.5 Data Storage*

Data storage for the acquired sensor data samples may be provided by the gateway. Implementations tend to vary from simple circular per-channel buffers to sophisticated distributed time-series databases with varying resolution.

In general, existence of local storage provides the flexibility to conserve network bandwidth and to reduce the load on cloud resources by selectively reporting data that meet specified conditions.

*4.1.6 Edge Analytics*

Edge-level analytics processing is becoming more common on more powerful gateways and fog nodes. These can be systems that monitor key indicators to detect failures of the attached equipment, such as pumps or motors, or predict them for informed and timely preventive maintenance. Advantages of edge-level analytics include proximity to data sources and the ability to process high-frequency data samples with low latency and to conserve network bandwidth by not sending them to the cloud. Edge analytics typically runs in real time and it can operate in disconnected mode.

Since AI and ML algorithms need large amounts of data to be successfully trained, edge analytics is usually developed to work in conjunction with its counterpart in the cloud. ML and AI algorithms, such as neural nets are developed in the cloud using its massive data aggregations and compute power. Portions of the developed algorithms, such as inference engines, that require fewer execution resources can be placed at the edge nodes. The two sides tend to work in tandem, with the edge executing algorithms locally to generate inferences and act upon them and/or report them to the cloud for further analysis and model improvements as appropriate.

## 5 IOT SYSTEM CLOUD COMPONENTS

This section describes cloud components that provide the IoT system functionality. They can operate in public or private clouds as an overlay over the "standard" cloud services such as virtualized compute and storage. This section covers core IoT functions involved in managing and processing data and control flows between the edge on one end and their delivery to cloud services and applications on the other end. Operational details of cloud applications and services - such as analytics, AI, and visualization – are not covered as they are beyond the scope of this document.

IoT-specific core cloud components include:

- Edge Interface and Data Ingestion
- Time-Series Stream Processing
- Data Aggregation and Storage
- Security and Management

Major functional blocks that commonly implement these functions are depicted in Figure 5.

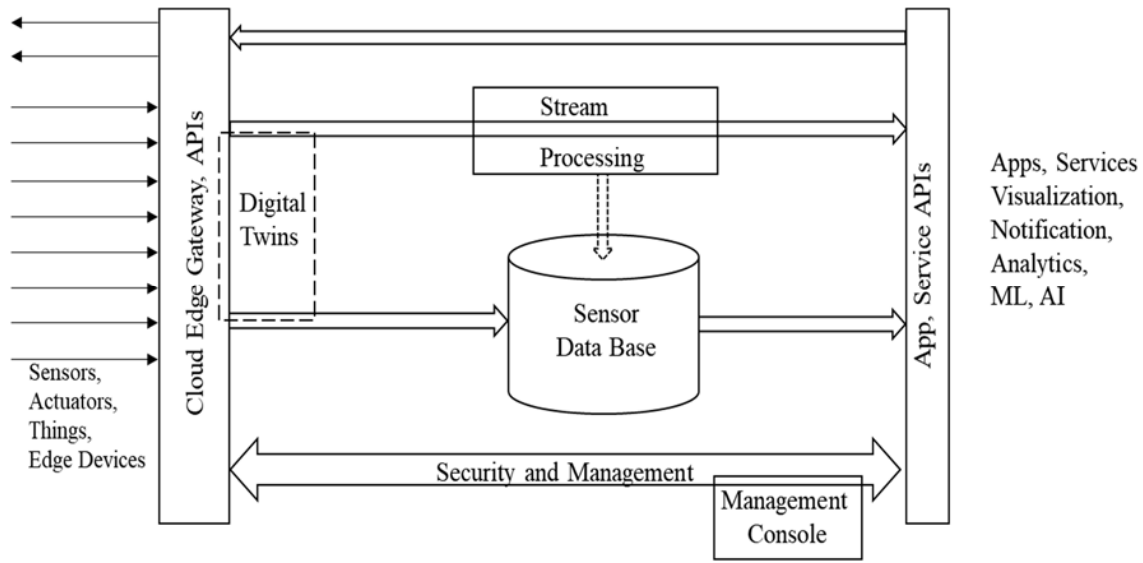

Figure 5 IoT Cloud Components

The box labeled cloud edge gateway is a functional module hosted in the cloud that interfaces with edge devices and things. It also acts as a security boundary and a control point between systems at the edge and the inner core components of the cloud.

Stream data processing and storage are shown as the two parallel paths that may supply data to the backend applications and services. Some of the incoming data may go through only one path commensurate with their processing needs, and some of the data may go through both paths to be processed both as an incoming stream and stored for archival and batch processing purposes. These two paths are sometimes referred as the data in motion, for streaming data, and the data at rest for stored items in the database that are used in the more time consuming and comprehensive types of processing, such as the analytics and ML model creation. This dual-path design is sometimes referred to as the lambda architecture, ostensibly because its split stream/storage data path resembles the Greek letter lambda, $\lambda$, (turned sideways).

Some IoT cloud systems implement digital twins as cyber representations and cloud replicas of endpoints. Their use and characteristics are described in a subsequent section.

Security and management is a cross-cutting system function that implements a control plane and can touch most of the other functional blocks. It is described in more detail in a later section.

Figure 5 also contains a block labeled Apps, Services APIs that represents cloud interfaces for the backend applications and services to access and retrieve IoT data. It may also represent a security boundary and access control point for service authentication and authorization. This is a conceptual representation and in many contemporary implementations cloud service interfaces tend to be bespoke APIs provided by individual applications, such as stream event processors and specific data bases.

## 5.1 Cloud-Edge Gateway and Data Ingestion

Cloud edge gateway is the boundary and a delineation point between the edge components and the cloud. Its functions are similar in many regards to an IoT gateway, with the key difference that it operates within the cloud security perimeter. The cloud-to-edge gateway or hub is the gating point between the external edge components with varying levels of security capabilities, and the highly secure and trusted cloud execution environment. It typically implements security checks such as access authorization, authentication, and possibly inspection as well as filtering of inbound data flows and messages based on their senders, content, and intended operations.

Edge connectivity provides access points for receiving data from the edge and for relaying actions to the edge components, such as the actuation commands and configuration of operational set points.

The primary functions of cloud data ingestion are to provide access points and methods for edge sources to deliver data to, and to provide the scaling necessary to handle the offered data volumes. Ingestion of input data primarily includes dealing with volume and routing to the appropriate destination.

Messages from the edge containing data reports can be delivered by simple HTTP posting to a configured ingestion URL where the receiving cloud logic can parse and route them as appropriate. Another common approach is to use a publish/subscribe mechanism. In those implementations data sources are regarded as publishers and they post their messages to preconfigured intermediate receiving rendezvous points. Applications and services in the cloud can receive data from the sources of interest by subscribing to their messages at the corresponding rendezvous points. Publish/subscribe matching points are usually implemented as brokers in the cloud. They manage active publishers and subscribers, and often provide some additional services such as persistence of sent messages until they are received by all active subscribers. The use of pub/sub mechanisms results in decoupling publishers and subscribers that enables asynchronous message delivery even in the presence of temporary unavailability or disconnectedness of its participants. It also allows messages to be received by multiple subscribers which simplifies management and saves edge bandwidth in systems where multiple services may need to process the incoming data. One of the most popular messaging pub/sub mechanisms in IoT systems is MQTT which is comparably lightweight and has a small implementation footprint.

The data volumes that cloud access points need to handle may be very high. Even when the individual data sources have relatively low frequency of reporting, aggregate data rates from a large number of endpoints can be considerable. In order to support high data ingestion rates, cloud access services may need to implement some form of scalability techniques, such as the creation of multiple instances of input servers or message brokers and balancing the load across them. A number of load balances and IP sprayers are available from the commercial providers and as open-source implementations. There are also specialized systems, such as Kafka, that can provide scalable data ingestion. Kafka is a high-throughput implementations of a durable messaging and publish-subscribe system that provides high reliability and scalability through portioning and replication. It was originally designed for handling web streams, such as the social network postings.

## 5.2 Digital Twins

The term digital twin generally refers to a synchronized cyber representation and replica of a physical thing that mirrors its states and behaviors. Depending on their implementation and levels of integrated models, digital twins may be used in all stages of a thing's lifecycle, including design and build phases in addition to operation.

In its basic form, an IoT digital twin can be a data structure in the cloud that represents physical things, such as sensors and devices. Twins receive data and status updates from their thing counterparts and thus mirror the thing states as close to the real time as possible.

IoT digital twins often have representation of the actual as well as of the desired states of the device. Applications and services may modify the desired state to cause the twin implementation to issue the necessary commands to the device to modify its state accordingly and to bring the two into compliance. Cloud-based applications and services can use and access digital twins as device proxies instead of accessing the endpoints directly. This mode of operation has several potential advantages:

- Faster access
- Always available
- Saves bandwidth and power
- Abstract representation and interfaces

The twin state representation is always available, even during the periods of time when the device is disconnected or sleeping. This allows delay-tolerant applications to proceed without having to wait for the temporarily inaccessible devices.

Savings of bandwidth at the edge and on the way to the cloud occur when access to digital twins can substitute for direct access to the devices. Bandwidth is usually in ample supply in the cloud where the digital twins reside, but it may be costly or restricted at the edge. Power may be saved in constrained environments by not forcing the device to wake up and to power up its transceiver to send a report. A cloud hosted digital twin can be accessed by the cloud applications on demand and without the latencies involved in accessing the actual device or waiting for it to report. Obviously, this access obtains the last reported thing state which is not necessarily the current one. That might be OK in many cases since instant synchrony in distributed is not possible anyway due to the inevitable communication delays. The target design point in such systems is to apply updates in a way that ensures that multiple copies eventually converge to the identical state in a consistent manner.

Abstract representation refers to the uniform data and command formats that digital twins can maintain for cloud applications to query thing data or to issue instructions to change its desired state. This approach can simplify implementation and increase portability of applications by hiding device differences, intricacies and details from the cloud applications. This is accomplished by using the twin implementations to translate between the generic representations and the device-specific data formats and commands.

In addition, IoT digital twins may be enhanced by incorporating device physical models and combining their outputs with the real data to perform simulations, analysis and predictions of thing and system behaviors.

### 5.3 Real-Time Stream Processing

Real-time stream processing operates on data in flight on its way from the origin to the consuming applications and services in the cloud. It captures the current state of the monitored IoT system. Various forms of streaming analytics can be deployed to detect system events and states that may indicate potential concerns or anomalies that need attention and may require immediate processing and reaction. This subsystem may also implement short-term storage to facilitate causal analysis of data that preceded events of interest.

In its simplest form, stream processing forwards events and alarms to their intended destination, such as the dashboard for visualization or direct operator alerting via messaging or email. More complex forms of real-time processing involve handling of events and execution of functions and transformations that can involve multiple input streams and produce multiple outputs. This can be implemented as a messaging or publish-subscribe pipeline where various functions or micro services take input from one or more sources, transform it, and produce one or more outputs. Those outputs can be forwarded to other functional transformations, thus enabling creation of complex pipelines and flows. Various output stages may also be delivered to other cloud services and applications via data-retrieval APIs.

Some implementations of the IoT streaming processing path include a form of storage that provides fast access to the detailed recent history of data reports and events. The idea is to keep the high-precision data around for a while to be able to roll back and analyze significant events by examining the preceding relevant state transitions that may have caused them. Afterwards, data can be aggregated and down sampled for trend analysis and transfer to the longer-term archival storage.

Short-term storage for streaming data is intended to provide the means to query data by time and to replay recent events for uses by streaming analytics and similar applications. It is usually optimized for fast access and its entries may be kept in memory or in the fast-access storage devices such as the solid-state devices (SSDs).

This type of storage has comparatively short data retention times, on the order of hours or days. Its implementations tend to resemble large circular message buffers. Due to the absence of the archival requirement, they have much less structure and overhead than traditional databases.

## 5.4 Cloud IoT Data Storage

IoT data and events ingested from the edge sources can be stored for batch processing and archival purposes such as the long-term comparisons or auditing. Computationally intensive applications, such as ML and AI, often work with large collections of data in the batch mode to create and refine inferencing models. Analytics, such as the building management, may use swaths of archival data such as the building telemetry on comparable days from the previous seasons and years to improve their optimizations and daily and hourly predictions.

Cloud implementations of IoT storage services typically consist of an input stage that receives incoming data through posting or subscriptions to the topics to be stored and writes them in the database. On the output side, the storage service implements support for queries and APIs for data retrieval. The central piece in this stage is the sensor database itself.

Properties of IoT Data

IoT data have some unique properties that impose additional requirements on their storage and retrieval. Some of the key ones include:

- Time-series data and metadata
- Semi-structured and unstructured data
- Lifecycle management, data retention
- Data-access patterns, writing and retrieval

The term time-series is usually applied to IoT data sampled and reported at regular intervals. This form of data acquisition can result in continuous streams of reports from the specific endpoints as data channels. Irregular events, such as changes of discrete states or exceptional conditions are reported as irregular events when they occur. IoT measurements and events are commonly marked with time stamps at capture or as close as possible to their point of origin. Time stamps are used for data correlation and access and need to be stored in the database. IoT data may also include additional metadata in the form of loosely structured key, value pairs that need to be stored and retrievable with or in conjunction with the related sensor readings and events.

In general, IoT data have characteristics of unstructured and semi-structured data in the sense that they have a great variety of formats and length of records, partly due to the variable sets of metadata. As such, they are generally not a good fit for the rigid schema of the structured data.

In terms of storage access patterns, incoming IoT data tend to be time-series sequences, thus their writes tend to be sequential and mostly appends. Depending on the nature of queries, retrievals can result in random reads of individual values or in larger sequences for data defined by the time intervals or attributes, such as proximity in terms of location or domain membership. If the database implementation provides some degree of control of its mode of data layout, it is generally advantageous to structure IoT data in storage in a manner that makes servicing of the common case of large queries efficient.

*5.4.1 Types of IoT Databases*

IoT data have high volumes and rates (throughput) and tend to appear in records of variable length and structure. They are often serialized as variable-length arrangements of key, value pairs of readings and metadata with time stamps.

Databases used for IoT storage generally belong to the category called NoSQL databases. NoSQL is a somewhat of an exclusionary definition meaning databases that do not use a System Query Language (SQL) that is common in the traditional relational database systems. Based on their structure and primary mode of operation, NoSQL databases tend to fall into the following major categories:

- Key, value stores
- Document-oriented databases
- Column-oriented databases
- Graph databases

Key, value stores or databases logically consist of stored values addressed by keys. Due to the nature of IoT data, the key, value and document-oriented databases are the most common types in IoT system implementations.

Document-oriented databases treat entries as documents that may be indexed by various keys. Documents in entries can be almost any text, with XML and JSON being quite common. In IoT applications, this allows for direct storage of input data that are often formatted as JSON strings. Documents can be of variable length which can be an additional benefit in IoT systems as new types of sensors with different message formats and lengths are introduced.

Column-oriented databases tend to be used mostly for massive applications such as the web-page index storage and processing, mostly batch, for search applications.

Graph databases use graph structures for semantic representation and queries of data. They consist of collections of nodes and edges, with the edges representing the relationships between nodes. Graph databases can be useful for representing complex structures and relationships in IoT systems, such as the layout and connections among the thousands of components in an HVAC system of a large building. Such representations can be used to visualize, model, analyze, and optimize complex system-level behaviors, such as the flows of energy and heating and cooling fluids in a building.

Cloud and web database systems can be very massive and span thousands of servers in multiple geographic regions. They often focus on capacity, scalability and fault tolerance by means of partitioning and replication. Cloud databases can be designed for structured or unstructured data with stream or batch processing modes and queries for data retrieval, and various models of consistency and convergence among the partitions and replicas of data.

## 6 CONTROL PLANE: SECURITY AND MANAGEMENT

The control plane in IoT systems consists mainly of security and management components and functions. It is in charge of keeping the data-plane part up and running securely to fulfil the overall system mission.

### 6.1 IoT System Security

IoT systems have some unique characteristics and security requirements that influence the selection of appropriate techniques and policies to deal with them. In terms of exposure, unlike most traditional IT systems, IoT systems can have direct impact on the physical world. When compromised, IoT systems can endanger the safety of people, environment, and production equipment and processes. In terms of topology, portions of IoT systems can reside in physically unprotected environments where they may be subject to tampering, eavesdropping, and spoofing.

Otherwise, like in any other complex IT system, security needs to be designed to maintain system confidentiality, integrity, and availability [9]. An IoT system security design usually starts with risk assessment and threat analysis that determine the levels and influence the type of techniques and methods to be implemented for its defense. Security design and implementation includes both static (built in) and dynamic (operational at runtime) components and activities.

Static design aspects include provisions for securing of nodes, data, and communications. Nodes at the exposed edge can be fitted with hardware assists to strengthen their security capability and posture. The common choices include trusted platform module (TPM) and variants of trusted execution environment (TEE). TPM is used to ensure that a node is securely booted with a trusted version of OS and possibly applications. TEE provides a separate trusted execution environment to store private encryption keys and materials, and optionally provide acceleration for more compute intensive forms of encryption and decryption of data for protection in both communication and storage. In this way, less capable edge nodes can be enabled to use the generally recommended but computationally demanding network encryption such as DTLS and TLS. On the more capable nodes, additional security may be provided by software isolation mechanisms, such as virtualization and use of containers.

A useful design technique in IoT systems is to deploy network segmentation. In this way, different protection levels can be implemented at different parts of the network, added security checks can be deployed when crossing segment boundaries – such as between the edge and the cloud – and parts of the network can be disconnected to isolate compromised nodes when the system is under attack.

While more of an operational than an architectural concern, it is generally a good practice to deploy defensive postures commensurate with the essentially closed nature of IoT systems to reduce their attack surfaces. These can include preference for certificate-based node authentication, coupled with disabling or disallowing password logins. Points of vulnerability, such as remote access, unsafe protocols like ftp, and even local shell should be blocked or removed. Communication should be directed only to a limited and well-known set of authorized system nodes and services and unsolicited connection requests should be blocked.

Software components of an IoT security system include the design and implementation of security agents that typically reside at each node to monitor activity, manage security policies and credentials, and communicate with the system's security management console. An edge security agent is depicted in Figure 4 and the security management console is depicted in Figure 3.

At runtime, when the system is in operation, node security agents and management console work in concert to maintain and update the system security posture and to actively monitor its state. Security agents locally apply and enforce relevant security policies, manage and refresh credentials, and actively monitor node behavior for anomalies that may indicate breaches or probing, such as unusual volumes of traffic or interaction attempts from or with unsanctioned parties.

The security management console combines the information obtained from nodes and their network neighbors, system logs, algorithmic and model predictions, and optionally global Internet monitoring sources, to detect breaches in real time and take the appropriate remedial actions when they occur. These actions need to be previously defined and designed as incident handling procedures that the operators can refer to when incidents are detected. Common actions include disconnecting the affected nodes to limit the impact of the incident until it is over. After the fact forensic analyses may be also applied to security monitoring data to discover the cause and nature of the breach and to devise system policies and procedure to guard against the similar ones in the future. Incident handling procedures should also indicate how to remediate the affected nodes and when and how to safely bring them back into the system.

## 6.2 IoT System Management

The management portion of the control plane is basically in charge of shepherding an IoT system by keeping its components functional and their software and firmware up to date. The management system is typically implemented as a collection of individual management agents installed on system nodes that communicate with the central management console. Its primary functions are: (1) provisioning of nodes, (2) registration and discovery, (3) firmware and software updates, (4) monitoring and management.

One of the key management system functions is to maintain a registry of system nodes. A registry can be a form of database that identifies the known and therefore authorized nodes. It can also perform additional services, such as act as a directory for node discovery, and to provide storage of node access and security credentials. Nodes not recorded in the registry should not be allowed to participate in IoT system operation. Nodes removed from the system should be deleted or marked as such in the registry.

The process of adding new nodes to the system is often referred to as commissioning and one of its key functions is the node provisioning necessary to complete the task. A new node needs to be assigned a unique system identifier and/or a name by which it can be referred to, addressed, and its reports tagged. In the process of commissioning, a node is registered, assigned security credentials, such as a digital certificate, and given system access points indicating where to direct its reports and queries. This step may also involve connection and interfacing to other enterprise systems - such as billing, CRM, or LOB – and to applicable external systems, such as a telco system for SIM provisioning.

Before a node becomes active and during its operation, the management system is in charge of updating its firmware and software. During normal operation, the management system continually monitors nodes for operational and configuration errors in order to take appropriate remedial actions when necessary. These may be affected through agent-supported administration and policy-based mechanisms via tools such as remote access and configuration.

## 7 IMPLEMENTATION CONSIDERATIONS

System designers and developers have to make a number of technical decisions when implementing IoT solutions for specific applications. This process starts with a definition of the purpose and intended benefit of the IoT system which in turn dictates choices of what to measure and the number and types of sensors to be installed for the purpose. Location and properties of the target installation area influence decisions on the type and topology of networks needed to connect them. End node design involves decisions such as the placement and capacity of hardware and functionality of software. Hardware spectrum of options covers a wide range from embedded microcontrollers to fully functional desktop or server CPUs with memory management and hardware support for virtualization, coupled with optional hardware assists for security, such as TPM and TEE, and performance accelerators such as GPUs, FPGAs and ASICs. The software spectrum of applications and runtime support can range from embedded firmware, all the way to hypervisors running multiple virtual machines with multiple containers. Special design provisions may need to be made in quite common cases where the equipment needs to be able to operate in harsh environments and in unattended modes with no human operators.

A major system design decision that needs to be made early on is the level and nature of security that will be required for nodes to be authenticated and authorized to communicate securely with other parts of the system. It has to be implemented by all access and interaction points on all nodes and networks in the system. Another early design decision should be the definition of the information model, and associated serialization mechanisms and transport protocols.

In the cloud, design choices include the ingestion capacity and corresponding tools, message routing rules and processing or a pub/sub system such as MQTT, event processing and stream analytics subsystems, sensor database for archival storage, such as one of NoSQL types, visualization dashboard and command/control console, ML and AI tools and whatever application choices may be relevant for the vertical application domain that the IoT system is supposed to serve. JSON and XML are commonly used for the data serialization of payloads. A data or information model needs to be adopted or defined for semantic annotation and machine-level interoperability between senders and receivers.

In terms of software implementation, it is generally beneficial to opt for cloud-first style of tools and practices that use microservices with well-defined APIs to support modularity and flexible placement of functions. The late binding principle should be followed in the sense that placement of system functions and processing modules can be deferred until installation or even runtime, as opposed to being fixed at the design time. In this way, edge vs. cloud function-placement decisions can be optimized for the requirements and system configuration at hand, and even readjusted to match the load variations when the system is in operation.

# 8 SELECT REFERENCE MATERIALS

# APPENDIX – IOT DATA MODELS AND INTEROPERABILITY

IoT systems have a requirement of machine-level semantic data interoperability which is not a required or generally implemented property of the worldwide web. In the common usage of www, end-to-end semantic interoperability is between humans (content creators on one end, and browsing readers on the other) and not machines. Thus, IoT system designers must devise a method for M2M semantic interoperability. This appendix outlines the common principles and practices used to accomplish that purpose.

Data sources and consuming services in an IoT system communicate via application-layer messages to exchange payloads containing items of interest, such as sensor data and commands. The payloads are often represented as "name" : "value" pairs encoded using JSON or XML. A message or report from the edge typically includes a unique identifier of the data endpoint as some combination of node identifier and of the originating sensor, followed by one or more attributes and their values – such as temperature reading and measurement units – and a reference time stamp. A portion of such message reporting temperature reading of 77.6 F may look something like this:

```
{"id":"150a3c6e-bef0e","temp":"n:77.6","unit":"0F","DateTime":"t:2020-07-15T14:50:07Z UTC"}
```

This machine-to-machine (M2M) communication requires machine-level semantic interoperability in the sense that the target node need to be able to "understand" the meaning of data that it has received. One way to accomplish that is for both communicating parties to use the same conceptual model of physical things that exist in their domain. In current practice, a common way to accomplish this is by having all parties use the same specification of an IoT information model.

**IoT Information Model**

An IoT information model is an abstract, formal representation of IoT thing types that often includes their properties, relationships, and operations that can be performed on them. A thing information model needs to specify what the thing is, what it can do, thing properties and their values, and ways to interact with it. Some specifications provide only a definition of thing types and their properties, and they tend to be referred to as data models.

One of the primary tasks of an IoT information model is to facilitate semantic interoperability, i.e. a shared understanding of what the data means. It is used by the IoT server to provide the compliant software abstraction of the thing, and by IoT clients to properly interpret it and to be able to process the data accordingly.

Following the Internet principles and practice, clients and servers exchange payloads and rely on an information-model specification for proper encoding and interpretation of the exchanged messages. The two endpoints are otherwise decoupled, they can be developed independently of each other and operate on different platforms and runtime environments. In this context, endpoint refers to software agents that are acquiring (producing) or processing (consuming) real world data.

**Structure of an IoT Information Model**

Software objects that represent and model IoT things are often structured in ways that follow the common practices and terminology in object-oriented (OO) programming. They are also referred to by other names, such as thing descriptions and smart objects. Since objects are modeling real-world things, there are some implied semantics defined by the intrinsic physical nature of the thing that is being modeled. For example, a temperature sensor is commonly understood to measure temperature of something based on its characteristics and placement, such as the ambient air temperature or water temperature in a heating/cooling pipe.

Table 1 illustrates general the form of object-oriented IoT information model representations as abstracted from a number of IoT standard definitions [6][14][15][16][19][20].

Table 1: Structure of IoT Information Model

| Element | Description |
| --- | --- |
| Object Type | Physical thing being modeled (device) |
| Properties, Attributes | Object attributes, data, metadata |
| Interactions | Ways to interact with object, actions, events |
| Links | To other objects, compositions, and collections |

IoT data models may specify only object types and properties of modeled things and leave out the definition of interactions and links.

*Object Type*

Within each specification, names of thing abstractions indicate the type of the physical thing or the phenomenon that is being modeled. This field is labeled as Object Type in Table 1 and is often modeled in object-oriented systems as a class. Object and class types directly correspond to the specific real-world things that are modeled by the specification. They can include a variety of sensors and devices, such as temperature, humidity, thermostat, refrigerator, security camera, current drawn, metered units (kWh, gals or liters) and many others. It is customary to use a single object-class definition representing a category of physical devices or things, such as a temperature sensor, and to create specific object instances of that class for each individual physical instantiation of a thing in a particular IoT system.

Information-model specifications and standards explicitly name and define each type of the physical entity that they model. Those names, in effect, constitute a vocabulary of types with their related definitions. A vocabulary may be a simple list of defined names resembling a dictionary. In some specifications, the naming tree may be structured as a as a hierarchy to form a taxonomy to indicate possible classification relationships among the objects, and to guide the proper placement of future objects as they are defined.

To facilitate unambiguous machine parsing, the vocabulary provides a precise definition of how terms are to be named and used when referring to a specific object class, such as temp or temperature. By using the type names exactly as defined in the vocabulary, communicating parties can establish an accurate correlation of references at both sending and receiving ends. This is a simple practical way of establishing semantic interoperability at the model level.

*Properties, Attributes*

Attributes generally say something about the object and its state, for example for a temperature sensor its current reading in the applicable units of measure, e.g. degrees C or F. IoT standards usually specify the exact format of the name fields and their meaning in the vocabulary. They also specify the data types for the values associated with the particular property and attribute name fields - such as numbers, integers, or strings. Values or attributes in thing abstraction representations can be read-only, such as temperature reading, or writeable for actuation, or both. They can also include metadata, such as what the sensor is measuring, e. g. air or water temperature, minimum and maximum possible values of reading, and the like.

*Interactions*

Some information models define the types of interactions that a modeled object supports. These can include interfaces and methods that implementations may support, such as the types of requests and responses for reading data and for issuing actuation commands. They can also include designations of protocols and formalisms, usually APIs, to interact with the thing. Device supported APIs usually include retrieval of attributes and values, such as sensor readings and metadata. They may also include ways to request machine-readable descriptions of the thing, its object type and characteristics, or to activate the built-in methods that operate on the object's internal structures and outputs, including actuations. Many specifications also define events, which are generally asynchronous signals emitted by the device to indicate some change of state under observation.

*Links*

Some sort of linking mechanism or representation is often included in IoT information models, primarily as a mechanism to point to items of interest or to form groupings of related objects. Groupings of interest may be formed to facilitate coordinated behaviors of multiple devices, such as home-automation things. They may also be used to reflect structural properties of the physical connections between devices as installed and configured in the field that may be of interest to applications, such as indicating which HVAC zone a temperature sensor and an air-handling unit belong to.

Composition via linking may also be used to describe composite objects that include some more primitive basic types already defined in the specification. For example, a smart thermostat object type may be constructed as a composition of a temperature sensor, set point for temperature regulation and scheduling, and actuators of the heating and cooling systems.

**Metadata Handling**

Metadata, also called tags, are used in IoT systems to provide contextual semantics of where and how data are gathered. They can annotate data to make it more useful for a variety of post-processing services and applications, such as analytics and device/asset management. Metadata annotations can provide enriched contextual information in the form of additional attributes of a thing's functionality, geographic location, manufacturer, serial number, and the like.

Standardized information and data models often specify some metadata as object properties or attributes. These are typically inherent properties of the things being modeled, such as the units of measure used by a temperature sensor when reporting its readings. Some IoT applications may benefit from or even require additional metadata, such as on the structural relationships among things that are determined later in the system lifecycle, e.g. at installation time. Those can be provided as carefully defined and consistently named additional attributes and tags [18].

For example, additional metadata may indicate that an air temperature sensor in a specific room is part of an HVAC zone whose ambient conditions are regulated by a named air-handling unit. This kind of information can be consumed by an application to programmatically determine which HVAC components to actuate in order to reduce or increase ambient temperature as dictated by the associated thermostat schedule or a user request. Without the metadata, knowledge of structural relations among nodes would have to be hard coded into an application and customized for each specific installation, resulting in implementations that are brittle, not portable, time consuming, and prone to errors.